\documentclass[twocolumn]{aastex631}
\usepackage{graphicx}
\usepackage{natbib}

\pagecolor{white}
\begin{document}

\title{Flux and Color of WISE 0855-0714}
\author{Edward Wright}
\affiliation{Department of Physics and Astronomy, University of California, Los Angeles}
\email{wright@astro.ucla.edu}

\author{Jack Foley}
\affiliation{Department of Physics and Astronomy, University of California, Los Angeles}

\begin{abstract}
WISE 0855-0714 is the coldest known brown dwarf, located 2.28 pc
from the solar system. Discovered by the Wide-Field Infrared Survey
Explorer (WISE) in 2014 \citep{Luhman_2014}, the object is of
interest to scientists because of its low temperature ($\approx270$
K), proximity to the solar system, small mass ($\sim3-10\: M_{J}$),
and high proper motion.  The first observations of W0855 
by WISE in 2010 are heavily contaminated by a background
source. With 10.5 years of observations following the NEOWISE
reactivation in 2013 \citep{mainzer/etal:2014}, we present a robust
analysis of W0855's flux and color unobstructed by this background
source.  We obtain W1 = 19.3 and W1-W2 = 5.4 magnitudes with an error
of 0.37 magnitudes.
\end{abstract}

\section{Introduction}

Brown dwarfs are substellar objects of a mass between 13 and 75
$M_{J}$ (where $M_{J} = 1.898 \times 10^{27}$ kg is the mass of
Jupiter), 
in which the mass
is so small that the inward compression by gravity is insufficient
to overcome the Coulomb barrier between two hydrogen nuclei,
preventing hydrogen nuclear fusion found in all Main-Sequence stars
\citep{Basri:2000}. While the most massive ($ > 65 \:M_{J}$) brown
dwarfs fuse lithium, the energy output of all brown dwarfs fall
well below that of Main-Sequence stars, and thus brown dwarfs appear
very dim. Primarily, brown dwarfs emit light in the infrared.

WISE J085510.74-071442.5, hereafter referred to as W0855, is unique
among brown dwarfs for many important reasons. Its mass falls between
3 and 10 $M_{J}$, and is thus classified as a sub-brown dwarf or a
planetary mass object \citep{Luhman_2023}. Despite its small size,
scientists still observe W0855 fairly easily due to its proximity
to Earth, at a distance of $7.43$ ly ($2.28$ pc) from the solar
system \citep{Kirkpatrick_2021}. This makes W0855 the fourth closest
star or brown dwarf system to the Sun, behind Alpha Centauri, Barnard's
Star, and WISE J104915.57-531906.1 = 2MASS J10491891-5319100
\citep{Luhman16}.  W0855 is also the coldest known brown dwarf, at
an effective temperature between 250-300 K, providing an incredible
source to study the chemistry and physics of extremely cool
astronomical objects. This makes it especially good as an analog
for high mass exoplanets, exhibiting similar characteristics while
being close enough for rigorous observation and study \citep{Rowland_2024}.

The Wide Field Infrared Survey Explorer (WISE), launched in 2009,
presented an incredible leap forward in infrared astronomy capabilities.
Its high sensitivity, large wavelength range, and wide field of
view allowed scientists to collect a cutting edge all-sky survey
in the mid-infrared.  WISE was launched to study astronomical objects
bright in the infrared, such as brown dwarfs, asteroids, and
Ultra-Luminous Infrared Galaxies (ULIRGS) \citep{wright/etal:2010}.
WISE simultaneously surveyed the sky in 4 bands, from W1 to W4,
centered at 3.4, 4.6, 12, and 22 microns respectively. The WISE
system, especially its data processing modules, were greatly enhanced
by the NEOWISE program in 2011, which specialized in detecting near
Earth objects \citep{Mainzer_2011}.  After consuming its cryogen
and losing access to the W3 and W4 bands, and entering a hibernation
starting 1 February 2011, WISE was reactivated to resume observations
of near Earth objects and other infrared phenomena, and began
collecting data again in December 2013 \citep{mainzer/etal:2014}.
WISE continued to make observations until 31 July 2024, when its
orbit became too unstable and the mission was officially ended.
After reactivation, WISE observed for a total of 10.5 years.

W0855 was discovered by Kevin Luhman in 2014, who detected the brown
dwarf from its very high proper motion compared to background stars
\citep{Luhman_2014}.  Luhman drew his data from the first period
of WISE observation from 7 January 2010 and 1 February 2011, and
noted a discrepancy between the measured W1 (3.4 $\mu m$) and W2
(4.6 $\mu m$) fluxes, attributing the difference to stationary
sources behind W0855 strong in the W1 band at the time of observation
\citep{Luhman_2014}. Comparisons with Spitzer Space Telescope
observations taken on 21 June 2013 and 20 January 2014 confirmed
that the W1 photometry of W0855 taken in 2010 and 2011 was dominated
by two background sources, rather than W0855 itself \citep{Luhman_2014}.
After reactivation, subsequent observations of W0855 (e.g.
\citet{Wright_2014}) confirmed that the 2010 observations of W0855
were contaminated by a background stationary source, making the
brown dwarf appear bluer in its W1-W2 color than it really is.

\begin{figure}[tb]
 \plotone{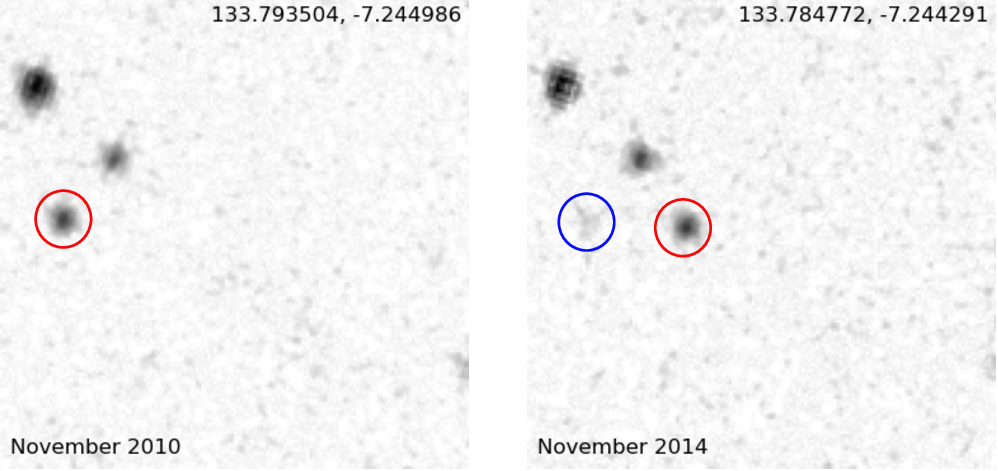} 
 \vspace{-1em}
  \caption{WISE/NEOWISE Coadder Image frames of W0855, taken in November 2010 (left) and November 2014 (right). 
  Right ascension and declination of W0855 in each observation is given in the top right corner of each image. 
  W0855 is circled in red, and the background source, revealed after W0855 moved between 2010 and 2014, is circled in blue.}
  \label{fig:beforeAfter}
\end{figure}

W0855 has been continually observed over the past 10 years, both
by WISE and other infrared telescopes,
such as Spitzer, and scientists have continued to sharpen its
physical and astrometric characteristics
(e.g. \citet{Kirkpatrick_2021}). The first spectrum of W0855 was
collected using the Gemini-North telescope and the Gemini Near
Infrared Spectrograph; these observations suggest a cloudy
atmosphere of water and water ice is present on W0855
\citep{Skemer_2016}. The James Webb Space Telescope's NIRSpec
spectrograph first observed W0855 in 2023, providing us the most
advanced spectral energy density, chemical makeup, temperature,
and stellar class data currently available \citep{Luhman_2023}.
Using this spectrum, \citet{Luhman_2023} arrived at an effective
temperature of 285 K and a mass between 3 - 10 $M_{J}$. Another
recent observation of its chemical composition, again using another
JWST NIRSpec spectrum, claims a detection of deuterium for the
first time in a substellar atmosphere outside of our solar system
\citep{Rowland_2024}.  This analysis by \citet{Rowland_2024}
presents an effective temperature of 264 K and a mass between
1.90 - 4.33 $M_{J}$.

\section{Methods}
\subsection{Astrometry}

To determine the most accurate photometric data for W0855, we first
pinpointed its astrometric parameters, namely its right ascension,
declination, proper motion, and parallax. To do so, observations of
W0855 were pulled from IRSA's WISE/NEOWISE Single Exposure Source Table for each
observational epoch, from which we collected W0855's right ascension
and declination. In total, data were collected for all 23 available
6 month epochs.  Assuming W0855's proper motion behaves linearly,
these collected positions served as the basis for a linear least
squares fit.

This model of W0855's motion needs to account for its initial
position, proper motion, and the annual parallax.  Its proximity
to Earth gives it a large parallactic motion of about 1 arcsecond
peak to peak.  However, the observed proper motion is much larger,
shifting 4 arcseconds per each observational epoch.  A 5 parameter
fit to the WISE positions of W0855 gives a parallax and proper
motion consistent with \citep{Kirkpatrick_2021}.  This 5 parameter
fit was used to determine the positions of W0855 at each observational
epoch free of photometric bias.  These coordinates were input
into the WISE single frame detection database maintained by the
NASA/IPAC Infrared Science Archive, providing the most accurate
photometric data of W0855 in each observational epoch.

\subsection{Flux}
In total, we used 4 different methods to determine the flux and color of W0855, 
providing the most robust photometric analysis of W0855 to date. 

\begin{figure}[h!]
 \plotone{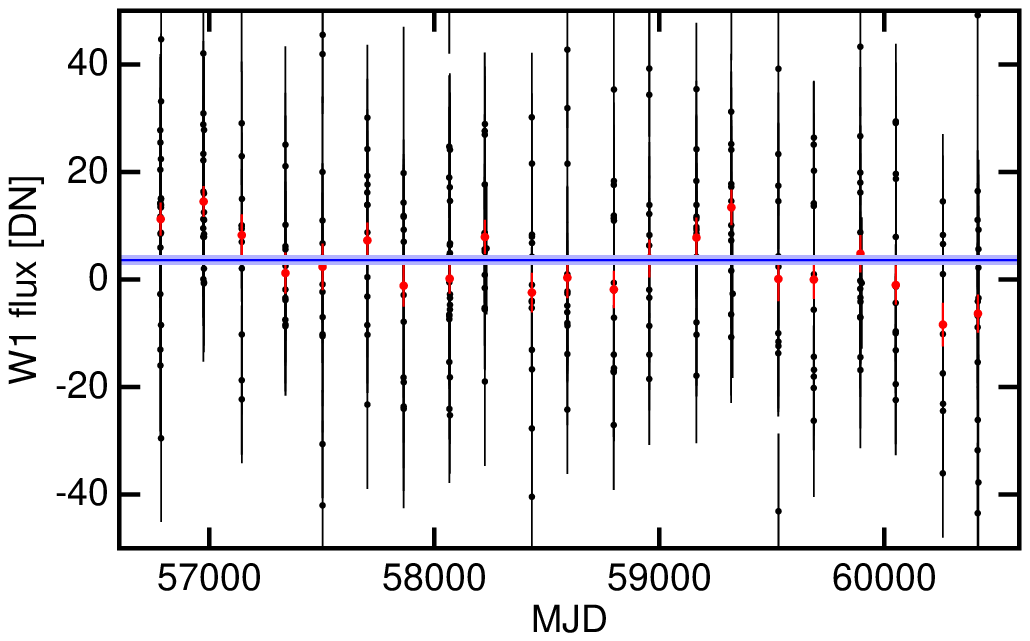}
 \caption{W1 Flux, in Data Numbers, of W0855 for each epoch, given in Modified Julian Dates. The red points and errorbars correspond to the epoch average. The black points and errorbars correspond to the measured W1 flux for each single frame W2 detection. The blue line and its margins correspond to the total calculated average W1 flux and error.}
 \label{fig:method1}
\end{figure}

Given the very low SNR in the W1 band, it is not reasonable to average the W1 magnitudes reported in the 
NEOWISE-R Single Exposure (L1b) Source Table available from IRSA. Since magnitudes are a logarithmic
measure of the flux, they are undefined for negative fluxes which are common at low SNR.  Even for positive 
fluxes, the magnitude is a non-linear function of the flux, so the average of the magnitudes is not the same
as the magnitude derived from the average flux.  For this reason we average using the ``w1flux'' and 
``w1sigflux'' values found in the long view of the database table.  These are reported in Data Numbers, or DN.
In the L1b images the pixel values are also in DN, but these are not identical to the integer values sent down
in the telemetry due to flat fielding. The flux in the database table is the sum of background-subtracted
pixel values over the image of a source. Then final conversion to a magnitude in the database is done using
\begin{equation}
m_{W1} = -2.5\log_{10}{F} + M_{0}
\label{equation:magConv}
\end{equation}
where $m_{W1}$ is the apparent magnitude in the W1 band and $M_{0}$ is the instrumental 
zero point magnitude calibration, corresponding to a source with a flux of 1 DN.  F is the flux 
of the source, taken as w1flux when the source has a signal-to-noise ratio $\geq 2$.
If this ratio is less than 2, F is taken as 
 $\max(\mbox{w1flux},0) + 2\,\mbox{w1sigflux}$, 
which presents an upper limit for sources with high error.
In principle the flux calibration given by $M_0$ could vary with time,
but in practice the calibration is very stable over the whole of the NEOWISE reactivation data.
We find $M_{0} = 20.762 \pm 0.003$ in W1 and $19.645 \pm 0.007$ in W2.
In the analysis reported below, we scale the W1 flux and W1 flux error values to the standard 
$M_{0}=20.5$ used for the coadded images.  Given the flux of 306.7 Jy for a
W1 = 0 star \citep{wright/etal:2010} this means 1 scaled DN in Figure \ref{fig:method1}
equals $306.7/10^{20.5/2.5} = 1.94 \:\mu\mbox{Jy}$ at 3.3526 $\mu$m for either a 
Vega-like spectrum or a constant $F_\lambda$ spectrum. 

Using the fitted coordinate positions found in the previous section, the raw flux density of W0855, 
in Data Numbers, was retrieved from the NEOWISE-R L1b source table in the W1 (3.4 $\mu m$) and 
W2 (4.6 $\mu m$) wavelength bands. 
Since W0855 has such a low SNR in the W1 band, the W1 flux density was collected for each 
instance of single frame detection of W0855 in the W2 band. 
A robust average of the W1 flux density was calculated for each epoch out of these single frame detections, 
and from these epoch averages we calculated a robust average over the entire observation period. 
These data are plotted in Figure \ref{fig:method1}. 

A second approach to estimating the W1 flux of W0855 is to fit all of the pixels taken in a given epoch
to a model that has a single point source at the position of W0855 determined from the W2 data, and
a spatially constant background that can be different for each frame.  This differs from the single frame
fluxes in that the single frame detections database primarily through the use of a single position derived from all
the W2 data for the epoch instead of the individual frame positions which are all slightly displaced from
the true position of W0855 by noise.  The pixel data is taken from the Level 1b frames available from IRSA.
One detail to watch out for is that the conversion from Data Numbers to fluxes can be different for each
frame.  This conversion is given by the MAGZP keyword in the L1B fits header, which gives the magnitude
corresponding to a source with a total flux of 1 DN. 

\begin{figure}[h]
\plotone{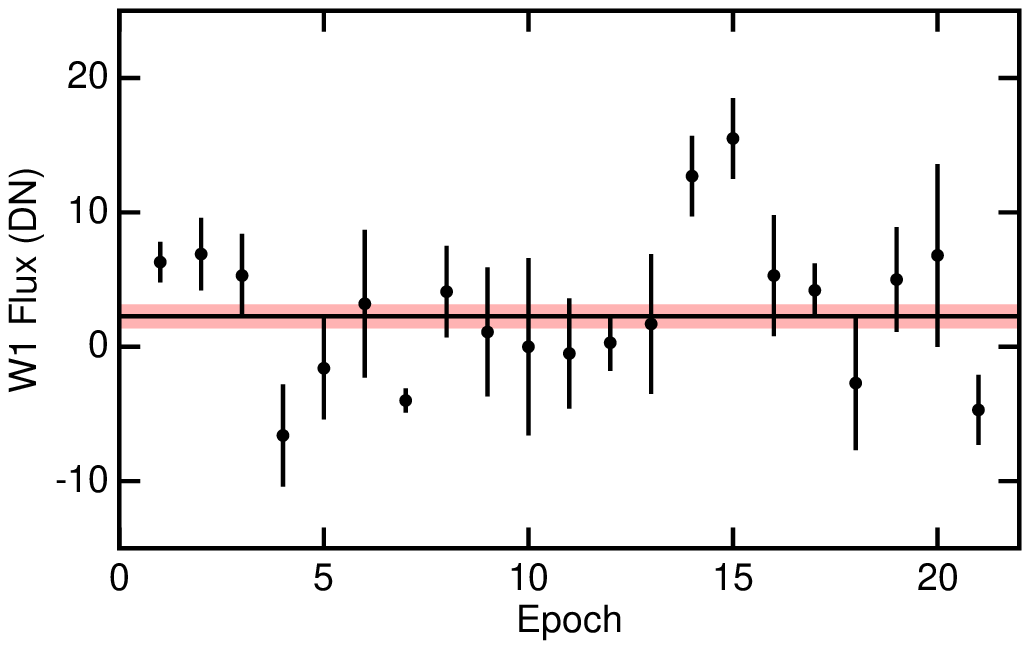}
\caption{W1 Flux, in Data Numbers, of W0855 for each epoch. Each point corresponds to the flux derived from the pixel values of the PSF centered at W0855's position, with the addition of a flat background, within 15" of the source for each epoch. The red line and its margins correspond to the total calculated average and its associated error.}
\label{fig:method2}
\end{figure}

Fitting the pixel values to the PSF requires a model for the PSF.  In this work we use a double 
Gaussian approximation to the PSF, with
\begin{eqnarray}
\psi(r) & = & \frac{1}{6.179}\left[
0.9841  
\exp\left(-0.5\left( r/2.539^{\prime\prime} \right)^2\right) \right.
\nonumber \\ 
& + & \left. 0.0159
 \exp\left(-0.5 \left( r/8.296^{\prime\prime} \right)^2\right)
\right]
\label{equation:psi}
\end{eqnarray}
where $r = \sqrt{\Delta\delta^2+ (\cos\delta \Delta\alpha)^2}$ is the radial separation. 
These data for each observational epoch are plotted in Figure \ref{fig:method2}.

\begin{figure}[h]
\plotone{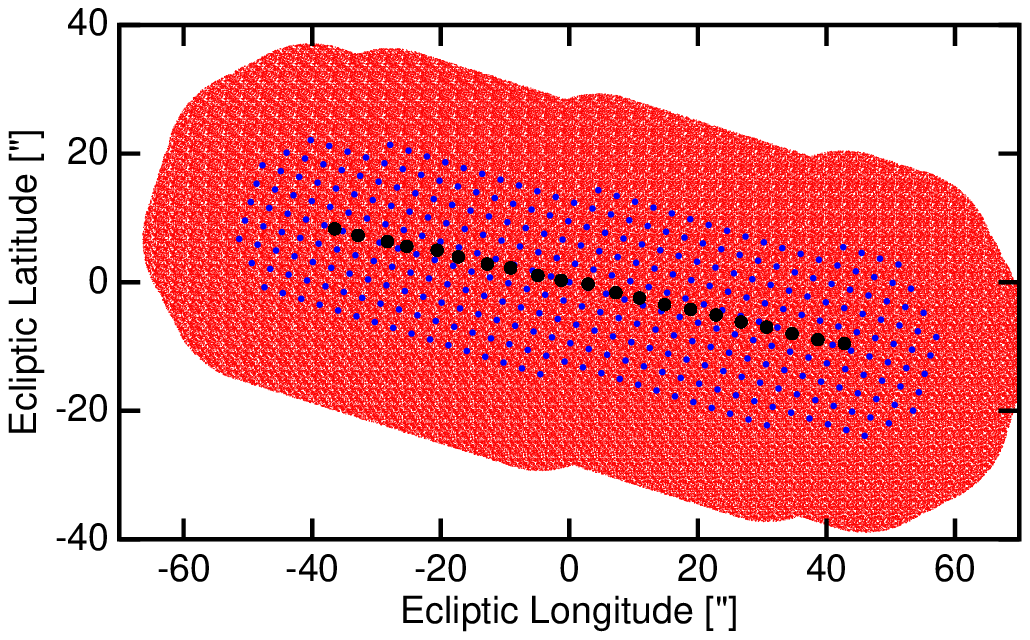}
\caption{All 300,048 pixels used to model the motion and flux of W0855 with an arbitrary background. The 21 post-Reactivation positions of W0855 are in black, the 349 background source flux pixels are in blue, and the centers of the source pixels are colored red. Each pixel is modeled as the sum of the source flux, times a centered PSF of W0855, plus any contributions from the background}
\label{fig:method3}
\end{figure}

Since the first two methods appear to have correlated residuals due to 
faint background stars, a third method was tried that allowed for an
arbitrary background.  An area of interest was defined as the union
of 15 arcsecond radius circles around the position of W0855 at each 
of the 21 epochs of reactivation observations.
Then a grid of background positions spaced by 2.75 arcseconds was set up
in that region.  This gave 349 positions.
Finally pixels that fell within 15 arcseconds of any background or source
position were extracted from the 327 frames that overlapped this region.
The final list contained 300,048 pixels.
Each pixel value was modeled as the sum of a source flux times a PSF centered
at the W0855 position at the time the frame was taken, plus a uniform
background for the frame, plus a sum of 349 background fluxes times a PSF
centered at each background position.  The final list of parameters
had 1 flux for W0855, 349 background fluxes, and 327 frame backgrounds,
yielding 677 total parameters.
Figure \ref{fig:method3} shows the 21 W0855 positions
in black, the 349 background flux positions in blue, and the pixel centers for
the 300,048 pixel values in red.
A least squares fit of the 300,048 pixels values by this 677 parameter
model was performed.  Since the least squares fit is not robust against
outliers, the 2.5\% of the pixel values with the highest residuals were
removed and the fit was redone.  This trimming process was repeated.
The final fit of 677 parameters to the remaining 95\% of the pixel values
gave a flux for W0855 in the W1 band of $2.78\pm0.78$ DN. 

\begin{figure}[tb]
 \plotone{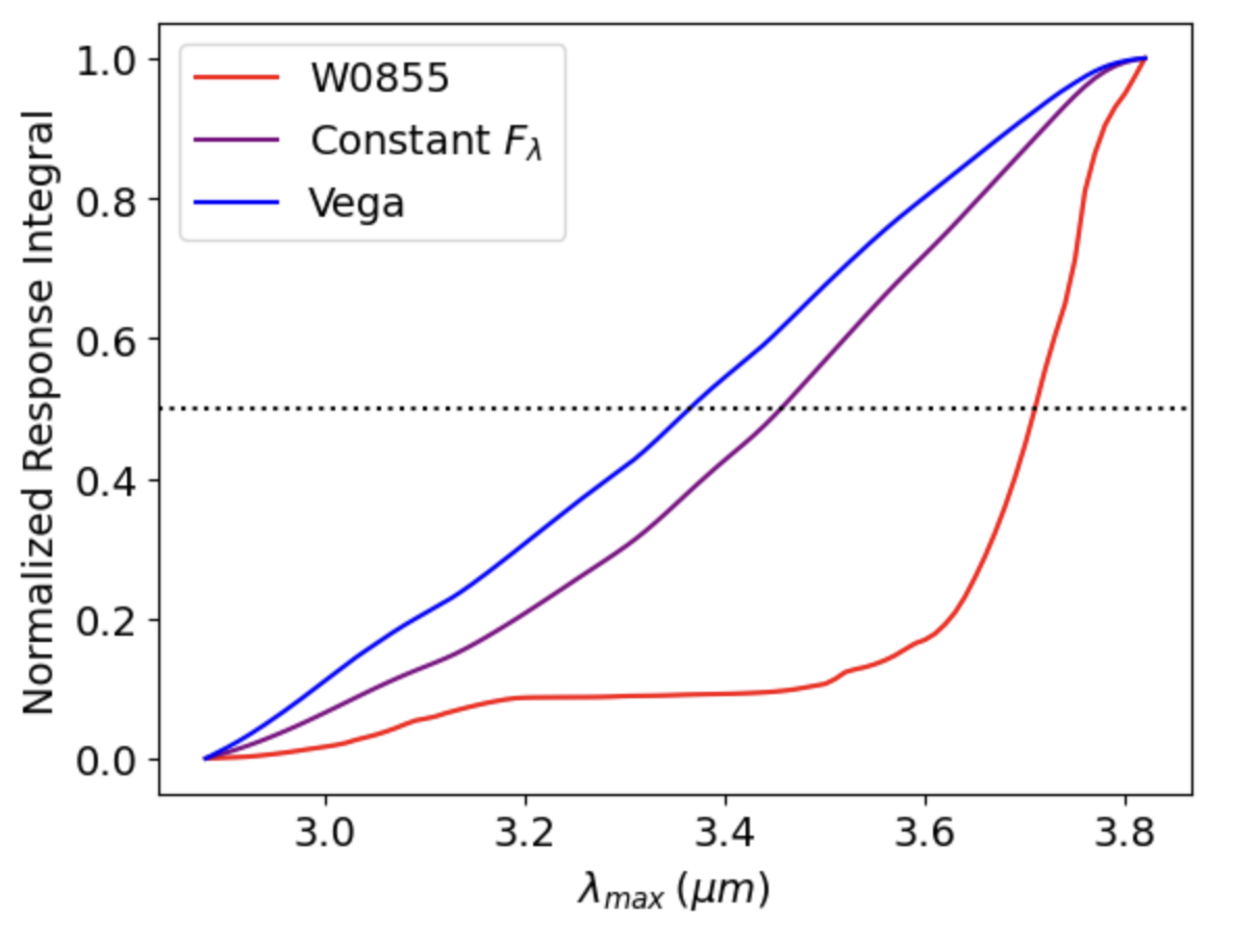}
 \caption{Normalized response integrals for W0855, Vega, and a constant $F_\lambda$. Each point corresponding to some $\lambda_{max}$ is the signal integral, Eqn \ref{equation:SignalIntegral}, integrated from 2.83 microns to $\lambda_{max}$, divided by the same integral integrated over the entire wavelength range for normalization.}
 \label{fig:method4}
\end{figure}

The infrared spectrum of W0855, as found by \cite{Luhman_2023}, provided another 
avenue to determine the most accurate flux density and color of W0855. 
The spectrum, given in erg/s/cm$^2/$\AA, was interpolated over a one micron range 
centered in the W1 band, between 2.83 and 3.83 microns, at an interval of 0.01 microns. 
To find the total flux over the W1 band, this interpolated spectrum, multiplied by the WISE
 relative spectral response function, was integrated over its entirety. 
 The relative spectral response function gives the instrumental response per received photon. 
 By dividing the JWST spectrum by the energy per photon at each wavelength ($E = hc/\lambda$), 
 we convert the spectrum into the number of photons per unit wavelength. 
 By multiplying this array by the relative spectral response function, and integrating across 
 our 1 $\mu$m range in the W1 band, we find the signal of W0855 in Data Numbers: 

\begin{equation}
S \propto \int R(\lambda)F_{\nu}d\ln{\nu}  \propto \int R(\lambda) \lambda F_{\lambda}d\lambda
\label{equation:SignalIntegral}
\end{equation}
where S is the signal, R is the instrumental response as a function of wavelength, 
$\nu$ is the frequency, $\lambda$ is the wavelength, and $F$ is the flux density, 
either dependent on wavelength or frequency \citep{wright/etal:2010}. 
To illustrate the redness of W0855, 
we calculated the response integral over a range from 2.83 $\mu m$ to some $\lambda_{max}$, 
again using Eqn \ref{equation:SignalIntegral}.  By varying $\lambda_{max}$ along a 1 micron range, we produced the curve shown in Figure \ref{fig:method4}. 
For normalization, and to clear up the proportionality, each of these values is divided by the same integral over the entire wavelength range.  
To compare, we repeated this process for reference spectra of Vega and a constant $F_\lambda$, which differ by $\approx 4$ powers of $\nu$. 
As shown in Figure \ref{fig:method4}, W0855 is far redder than either reference spectrum.

\subsection{Color}

The color of W0855, as detected by WISE, immediately follows from our calculations of W0855's W1 band flux. 
Our determination of color comes from the subtraction of the W2 Vega magnitude of W0855 from its W1 Vega magnitude. 
The logarithmic relationship between flux ratios and magnitude differences tells us that the color of W0855 derived in this manner is simply the ratio of W2 to W1 flux. Since W0855 is fairly bright in the W2 band, the W2 Vega magnitudes laid out by the IRSA catalogue service are sufficiently rigorous. 
Our prescribed methods produce  This is done so using  Eqn \ref{equation:magConv}.
As discussed later, W0855's spectral index diverges greatly from that of an A0 star, making it difficult to convert to Janskys with any validity; 
any straightforward conversions offered by WISE only apply to sources similar to Vega, with a spectral index close to 2 ($F \propto \nu^{2}$).  
As such, this paper makes its determinations of flux and color of W0855 using Data Numbers, 
which are not dependent on the spectral index of the observed body. All references to the received signal of W0855 are done in Data Numbers, and conversions are again derived from Eqn \ref{equation:SignalIntegral}.

\section{Results}

The first method, utilizing the W1 fluxes from all W2 single frame detections produced a total average value of  $3.17 \pm 0.9 $ DN after
scaling to a magnitude zeropoint of 20.5,
This value is the highest of all four methods. However, $\chi^2 = 58.4$ with 20 degrees of freedom indicating that the scatter from
epoch to epoch is larger than the expected scatter based on the scatter within each epoch.  

The second method, compiling the product of fluxes and the WISE Point Spread Function, plus a frame by frame flat background, produced a total average value of $2.95 \pm 0.91$ DN. The $\chi^2$ test performed on this data produces a reasonable ratio to the dataset's degrees of freedom. However, there is likely some contamination from faint background sources nonetheless.

The third method, which simultaneously fit for a 336 point spatially varying background plus a moving source, produced a total average value of $2.59 \pm 0.73$ DN.

The fourth method, integrating the infrared spectrum found by JWST, produced a total average value of $1.91 \pm 0.03 $ DN. 
This value is the smallest of all four methods. 
To illustrate how strongly W0855's flux varies at the red end of the W1 spectral response function, 
the signal curves in Figure 4 were converted into a logarithmic space to derive their spectral indices. 
We determined that a $F_\nu \propto \nu^{-14}$ power law is approximately as red as W0855's spectrum
in the W1 band,
showing the extreme dependence of W0855's flux on frequency. 

\subsection{Color}

The color of W0855 follows immediately from the determination of its W1 flux. 
The robust average of W2 magnitudes, taken from the catalog results of all non-contaminated epochs, is 13.91 magnitudes. 
Taking an average flux from the WISE-based methods 2 and 3, and using Eqn \ref{equation:magConv}, the W1 magnitude is $19.27 \pm 0.37$ magnitudes. 
We discard the flux measured in method one due to its high $\chi^2$ result. As such, the W1-W2 color of W0855 is $5.36 \pm 0.4$ magnitudes.

\section{Discussion}
Interestingly, the color we derived in the previous section diverges greatly from that derived from 
Spitzer Space Telescope observations. 
\citet{Luhman_2014} report the [3.6] - [4.5] color of W0855  is 3.55 magnitudes. Clearly, 
W0855 appears far less red in Spitzer photometry than in WISE photometry presented in this work. 
This may primarily be attributed to the wider wavelength separation of WISE's W1 and W2 bands, 
centered at 3.4 and 4.6 microns respectively, compared to the Spitzer Channel 1 and 2 centers 
at 3.6 and 4.5 microns respectively \citep{Fazio_2004}. 

Our understanding of the nature and origin of W0855 is limited by its singular nature.
But we have successfully measured its W1 flux at a level 1.6 magnitudes lower than the 
90\% completeness limit of the CatWISE2020 catalog \citep{Marocco_2021},
and if it is possible to search (not just measure) in W2 to greater depth then
objects like W0855 but fainter due to greater distance or greater age could be found.
An obvious first step would be to redo the CatWISE2020 analysis on the entire WISE
and NEOWISE dataset.  This ``CatWISE2024'' would approximately double the amount of
data and give much improved proper motions.
The CatWISE2020 catalog is clearly limited by confusion noise in W1, but 
less confused in W2, so CatWISE2024 would also gain insensitivity.
The confusion noise is
reduced for faster moving sources with proper motions $> 2$ arcsec/year, but
these fast movers will also be trailed in the multi-epoch stack which reduces
the CatWISE2024 sensitivity. 
So we consider whether the methods
we used for the W1 band can be applied in the W2 band to search a much greater 
volume for more very cold objects like W0855.
Method 1 requires that the source be bright enough to be reliably detected in single frames at least one band,
and can not be used to search for brown dwarfs that are faint in W2 and fainter still in other bands.
Methods 2 and 3 require knowledge of the source position at each epoch which is not available
when searching for new objects.
However, synthetic tracking \citep{Shao2014} could be a promising approach for 
detecting high proper motion brown dwarfs.
In this method, frames are shifted and stacked using a grid of assumed right ascension 
and declination proper motions.  The WISE plus NEOWISE dataset spans 14 years, and
if we do the stacking on a grid of proper motions with a spacing of 0.28 arcsec/year, the
trailing using the closest proper motion will be smaller than $2$ arcsec which is well
under the 6 arcsec WISE FWHM.  
Covering a range of $\pm 14$ arcsec/yr in two dimensions, there will
be $10^4$ stacked frames to search for potential targets.  This is clearly a massive computational task,
but it can be reduced by an order of magnitude by using the 
single epoch unWISE coadds \citep{Lang_2014,Meisner_2017}
instead of the individual WISE frames.  
These coadds were used in constructing the CatWISE catalogs.

Synthetic tracking is just shifting and stacking on an assumed object motion.  It does
not attempt to fit and remove the static sky.  One can make a first approximation
to the least-squares fitting approach of our third method by constructing a grand
coadd of the static sky and then subtracting it from each epoch before shifting and
stacking the epochs.

With synthetic tracking one can probably search down to W2 = 17.5 magnitudes.
This is 3.6 magnitudes or a bit more than a factor of 25 fainter than W0855.  This
would open a search volume 125 times larger than the volume of the sphere centered 
on the Sun with radius equal to the distance of W0855.
One could also search for less luminous objects that are colder than W0855.
These could be older objects with the same mass as W0855 which is about 1 Gyr old
for a mass of 3 $M_J$.
These stacked frames might reveal objects with luminosities 4 times lower than W0855, 
and thus about 4 times older. 
Taking 250 K for the effective temperature of W0855, 
a 4 times lower luminosity means an effective temperature of ~180 K.
Equation 4 of \citet{wright/etal:2010} says that objects with an effective temperatures of 200 and 180 K
would 5 and 14 times dimmer in the W2 band than a 250 K object.
It may be necessary to loop over parallax leading to a full 5 dimensional search for sources,
but the parallax grid can be very coarse.  Any object with a parallax $< 1$ arcsec will
suffer minimal trailing losses in a search that  assumes no parallax, because the 
WISE PSF has a $\sigma \approx 2.5$ arcsec.

\section{Acknowledgements}

This is work is based on observations made by the NASA/JPL Wide Field Infrared Survey Explorer 
and Near Earth Object Wide Field Infrared Survey Explorer pipeline. 
This research has made use of the NASA/IPAC Infrared Science Archive (IRSA), 
which is funded by the National Aeronautics and Space Administration 
and operated by the California Institute of Technology. 
All tabular catalogue data of W0855 can be accessed via the IRSA Catalog service via \doi{10.26131/IRSA144}. 
All image frames of W0855 were accessed through the WISE/NEOWISE Coadder, and can be accessed 
via \doi{0.26131/IRSA535}. 
This work also acknowledges the JPL Solar System Dynamics Group and Dr. Jon Giorgini for the 
Horizons Ephemerides System, from which Earth's positional data used in calculating W0855's parallax were accessed. 
Further information on the Horizons Ephemerides System can be found 
at \href{https://ssd.jpl.nasa.gov/horizons/manual.html#ack}{https://ssd.jpl.nasa.gov/horizons/manual.html}.

\clearpage
\bibliographystyle{apj}

\end{document}